\documentclass{document}

\begin{document}

  \title{Gauge Models in $D$ Dimensions}

  \author{Douglas Moore}
  \email[~~]{douglas\_moore1@baylor.edu}
  \affiliation{EUCOS-CASPER, Department of Physics, Baylor University,\\ Waco, TX 76978, USA}

  \author{Jared Greenwald}
  \email[~~]{jared\_greenwald@baylor.edu}
  \affiliation{EUCOS-CASPER, Department of Physics, Baylor University,\\ Waco, TX 76978, USA}

  \author{Gerald Cleaver}
  \email[~~]{gerald\_cleaver@baylor.edu}
  \affiliation{EUCOS-CASPER, Department of Physics, Baylor University,\\ Waco, TX 76978, USA}

  \preprint{CASPER-13-1}
  \preprint{BU-HEPP-12-01}

  \begin{abstract}
    Utilizing the Gauge Framework, software under development at Baylor University, we explicitly construct all layer $1$ weakly coupled free fermionic heterotic string (WCFFHS) gauge models up to order $32$ in four to ten large spacetime dimensions.
    These gauge models are well suited to large scale systematic surveys and, while they offer little phenomenologically, are useful for understanding the structure of the WCFFHS region of the string landscape.
    Herein we present the gauge groups statistics for this swath of the landscape for both supersymmetric and non-supersymmetric models.
  \end{abstract}

  \maketitle

  \renewcommand{\thefootnote}{\alph{footnote}}


    \section{Introduction}\label{introduction}
  It is well known that the number of possible string derived models is on the order of $10^{500}$ \cite{Bousso:2000, Ashok:2003}.
  Consequently, any efforts to explore this landscape of string vacua require the use of high-performance computing and a choice of construction method.
  Each such method has access to different, often overlapping, regimes of the landscape; here we will focus on the weakly coupled free fermionic heterotic string (WCFFHS) construction formalism \cite{Antoniadis:1986, Antoniadis:1987, Kawai:1986_2, Kawai:1987}.
  The WCFFHS formalism has produced some of the most phenomenologically viable models to date, \cite{Cleaver:1999, Lopez:1992, Faraggi:1989, Faraggi:1992, Antoniadis:1990, Leontaris:1999, Faraggi:1991, Faraggi:1992_2, Faraggi:1992_3, Faraggi:1991_2, Faraggi:1991_3, Faraggi:1995, Faraggi:1996, Cleaver:1997, Cleaver:1997_2, Cleaver:1997_3, Cleaver:1998, Cleaver:1998_2, Cleaver:1998_3, Cleaver:1998_4, Cleaver:1998_5, Cleaver:1999_2, Cleaver:1999_3, Cleaver:1999_4, Cleaver:2000, Cleaver:2000_2, Cleaver:2001, Cleaver:2001_2, Cleaver:2002, Cleaver:2002_2, Cleaver:2002_3, Perkins:2003, Perkins:2005, Cleaver:2008, Greenwald:2009, Cleaver:2011} and is ideal for computer construction.
  Random examinations of the landscape, using this formalism, have been performed in the past, \cite{Dienes:2006, Dienes:2007_2}; however, due to the many-to-one nature of this construction a random survey of the input parameters has many endemic problems that are non-trivial to address \cite{Dienes:2007}.
  One way to approach these problems is to {\it systematically} survey the valid input parameters.
  Such systematic surveys have been performed in four large spacetime dimensions, \cite{Faraggi:2004, Assel:2010, Rizos:2011, Renner:2011b, Renner:2011c, Moore:2011}.
  Further, by restricting the class of models under investigation we can make the problem more tractable while still providing insight into the structure and ``texture" of the landscape.
  Herein we focus on such a restricted class, termed ``gauge models", first introduced and analyzed in \cite{Moore:2011}, and extend the results from four large spacetime dimensions to ten large spacetime dimensions.
  That is, we explicitly construct all layer $1$\footnote{We follow the convention that the ${\bf 1}$ and ${\bf S}$ sectors are not included in the number of layers.} WCFFHS gauge models from order $2$ through $32$ with up to six compactified dimensions.

  We begin with a short review of WCFFHS model building, subsection \ref{subsection:wcffhs-model-building}, and gauge models, subsection \ref{subsection:gauge-models}.
  Given the construction of all such models, one can extend the analysis found in \cite{Moore:2011}, expressing the occurence of various gauge group combinations in the appendix.

  \subsection{WCFFHS Model Building}\label{subsection:wcffhs-model-building}
    Within the free fermionic framework two inputs are required: the set of basis vectors, $\bm{A}$, and the GSO projection coefficient matrix, $\bm{k}$.
    In order to systematically build these models we need to systematically build the input set $\{\bm{A},\bm{k}\}$ ensuring that all of the modular invariance constraints are met.

    In $D$ large spacetime dimensions, the basis vector set is defined as
    \begin{equation}\label{eqn:1}
      \bm{A} = \left\{\vec{\alpha}_i \left|\right. \vec{\alpha}_i \in \mathbb{Q}^{(80-4D)} \cap \left(-1,1\right]^{(80-4D)} \right\},
    \end{equation}
    where $i = 0,1,\dots,L+1$.
    For our purposes we will always take $L$, the layer, to be $1$.
    Each of these basis vectors represents the boundary conditions of real worldsheet fermion degrees of freedom.
    We will be taking $\alpha^j_i$ with \mbox{$j = 0,\dots,(27-2D)$} to represent the boundary conditions of the left-moving supersymmetric string and $\alpha^j_i$ with \mbox{$j = (28-2D),..,(80-4D)$} to represent the right-moving bosonic string boundary conditions.
    The order of each basis vector, $N_i$, is the smallest positive integer such that
    \begin{equation}\label{eqn:2}
      N_i ~ \alpha^j_i = 0 \pmod{2}.
    \end{equation}
    Of course, the choices of these basis vectors are constrained by modular invariance in such a way that
    \begin{equation}\label{eqn:3}
      N_i ~ \vec{\alpha}_i^2 =
      \begin{cases}
        0 \imod{16} & \textrm{if } N_i \textrm{ even}\\
        0 \imod{8} & \textrm{if } N_i \textrm{ odd}
      \end{cases}
    \end{equation}
    and
    \begin{equation}\label{eqn:4}
      N_{ij} ~ \vec{\alpha}_i \cdot \vec{\alpha}_j = 0 \imod{8},
    \end{equation}
    where $N_{ij}\equiv\textrm{LCM}(N_i,N_j)$.

    Since we are dealing with $L=1$, we have three basis vectors, two of which will always be the same for every basis vector set we generate:
    \begin{itemize}
      \item The first basis vector, denoted $\mathds{1}$, contains the all-periodic boundary conditions: $(\vec{1}^{(28-2D)} ~||~ \vec{1}^{(52-2D)})$.
      \item The second basis vector is the SUSY generator, $\bm{S}$, which is $(\vec{1}^8~\vec{0}^{(20-2D)}||~ \vec{0}^{(52-2D)})$.
    \end{itemize}

    It is important to note that our definition of the SUSY generator is not formally accurate.
    Specifically, for even values of $D$ our definition generates $SU_2^6$ supersymmetry; however, for odd values of $D$ the WCFFHS formalism has difficulty expressing SUSY.
    Discussion of our treatment of SUSY is expanded upon following the presentation of the GSO projection.

    Once a modular invariant set of basis vectors has been generated, the states can be built.
    To do so we begin by generating all sectors as linear combinations of the basis vectors with $m^i_j \in \mathbb{N}$ and $m^i_j < N_j$, namely
    \begin{equation}\label{eqn:5}
      \vec{V}^j = \sum_i m^j_i\vec{\alpha}_i.
    \end{equation}

    The sectors describe how the worldsheet fermions, $f_j$, transform around non-contractible loops on the worldsheet
    \begin{equation}\label{eqn:6}
      f_j \longrightarrow  - \mathit{e}^{\mathit{i}\pi V^i_j} f_j.
    \end{equation}

    To each of these sectors we apply fermion number operators, $\vec{F}^i$,
    \begin{equation}\label{eqn:7}
      \vec{Q}^i = \frac{1}{2} \vec{V}^i + \vec{F}^i
    \end{equation}
    to build the charges. We can then express the masses of the states in terms of the charges as
    \begin{subequations}
      \begin{equation}
        \alpha' m_{left}^2 = \frac{1}{2}\left(\vec{Q}^i_{left}\right)^2 - 1
      \end{equation}
      \begin{equation}
        \alpha' m_{right}^2 = \frac{1}{2}\left(\vec{Q}^i_{right}\right)^2 - 2
      \end{equation}
    \end{subequations}
    Because we are working at the string scale and only interested in the low-energy effective theory, these states must be massless.
    Thus,
    \begin{subequations}
      \begin{equation}
        \left(\vec{Q}^i_{left}\right)^2 = 2
      \end{equation}
      \begin{equation}
        \left(\vec{Q}^i_{right}\right)^2 = 4
      \end{equation}
    \end{subequations}
    in a real fermion basis.

    However, once the states are constructed we must ensure that they are, in fact, physical.
    This requires the application of a GSO projection, and hence the specification of GSO projection matrix, $\bm{k}$.
    This matrix is, in our case, $(L+2) \times (L+2)$ and is constrained by modular invariance:
    \begin{subequations}\label{eqn:10}
      \begin{equation}\label{eqn:10a}
        k_{ij} + k_{ji} = \frac{1}{2}\vec{\alpha}_i\cdot\vec{\alpha}_j \imod{2}
      \end{equation}
      and
      \begin{equation}\label{eqn:10b}
        k_{ii} + k_{i0} = \frac{1}{4}\vec{\alpha}_i\cdot\vec{\alpha}_i - s_i \imod{2},
      \end{equation}
    \end{subequations}
    with
    \begin{equation}\label{eqn:11}
      N_jk_{ij} = 0 \imod{2}.
    \end{equation}

    It is clear from \autoref{eqn:10} that, in general, we have $\frac{1}{2}(L+1)(L+2) + 1$ degrees of freedom in our choice of $\bm{k}$.
    However, one of the degrees of freedom, our choice of the $k_{00}$ element, has no effect on the model generated so we fix it to $1$.
    This reduces us to $\frac{1}{2}(L+1)(L+2)$, meaning we can specify the lower-triangle of our GSO projection matrix.
    There is however, a caveat; not every choice of the lower-triangle yields a modular invariant matrix.
    This was shown in \cite{Moore:2011} and so will not be repeated here.

    Once a GSO projection matrix has been specified, the GSO projection can be applied,
    \begin{equation}
      \vec{\alpha}_i \cdot \vec{Q}^j = \sum_{l = 0}^{L+1} m^j_l k_{il} + s_i \imod{2}
    \end{equation}
    with $s_i$ being the space-time component of $\vec{\alpha}_i$.
    Now we are in a position to consider the space-time supersymmetry of these models.

    It is important to note that a subtlety arises in this survey when classifying models by number of spacetime supersymmetries; specifically, the maximum number of SUSYs allowed for a given model depends on the number of compactified dimensions.
    In order to simplify classification we will refer to models with the maximum number of SUSYs in a given number of compact dimensions as ${\cal N}_{max}$.
    Additionally, supersymmetry for odd values of $D$ is not well understood in the context of WCFFHS models.
    In such a situation we will consider models constructed from the all-periodic sector and a single bosonic sector, as before, and a third sector with a form analogous to the SUSY sector, though not strictly such.
    Videlicet, this does not generate supersymmetry, but it does have a congruous affect on the gauge groups.
    In all $L=1$ models, the GSO projection differentiates a ``model" into its ${\cal N}_{max}$ or ${\cal N}=0$ form as described in \cite{Moore:2011}.
    From this we can consider models in odd $D$ that have an agnate form to a supersymmetric model based solely on the GSO projection.

  \subsection{Gauge Models}\label{subsection:gauge-models}
    The Gauge Framework focuses on the construction of gauge models.
    These models are, in many ways, some of the simplest models that one can build.
    We can think of them as the basis from which more complex models can be built which makes them interesting as a starting point for systematic surveys.
    We can use what we learn about these models to guide further searches.
    Further discussion requires a more concrete definition of what a ``gauge model'' is.

    \begin{center}
      \parbox{.8\columnwidth}{
        \begin{definition}[Gauge Model]
          A model is a \textbf{gauge model} if it can be built from a set of basis vectors in which every basis vector beyond the all-periodic and SUSY basis vectors is bosonic of the form $(\vec{0}^{28-2D}~||~\vec{\alpha})$, within the free fermionic construction \textrm{\cite{Antoniadis:1986, Antoniadis:1987, Kawai:1986_2, Kawai:1987}}.
        \end{definition}
      }
    \end{center}

    In addition to the structural form described, we restrict ourselves to models with no left-right pairing allowing us to work in a complex basis.
    This choice allows for a few modifications to the expressions previously detailed.

    In particular modular invariance reduces for the gauge basis vectors because the left-movers become order $1$,
    \begin{equation}\label{eqn:3}
      N_i ~ \vec{\alpha}_i^2 =
      \begin{cases}
        0 \imod{8} & \textrm{if } N_i \textrm{ even}\\
        0 \imod{4} & \textrm{if } N_i \textrm{ odd}.
      \end{cases}
    \end{equation}

    Further, for the same reason, our masslessness constraints reduce to
    \begin{subequations}
      \begin{equation}
        \left(\vec{Q}^i_{left}\right)^2 = 1
      \end{equation}
      \begin{equation}
        \left(\vec{Q}^i_{right}\right)^2 = 2.
      \end{equation}
    \end{subequations}

    This, coupled with the form of the basis vectors, requires that the only massless states are either vector bosons or arise from a sector including the SUSY sector.
    Thus, when considering GSO projection of the gauge states we can simplify our GSO projection as well
    \begin{equation}
      \vec{\alpha}_i \cdot \vec{Q}^j_{right} = \sum_{l = 0}^{L+1} m^j_l k_{il} \imod{2}.
    \end{equation}

    These simplifications vastly improve the efficiency of the survey process without loss of too much generality.
    In fact, the Gauge Framework has improved upon the performance of its predecessors by a factor approaching $10^4$.
    We can now consider the statistics of the unique models generated at layer $1$.
    \section{Statistics}
  In conformation with \cite{Renner:2011a}, we find two $\NMax$ and six $\NZero$\footnote{An additional model can be produced utilizing chiral isings, however such models are not considered here.} models in $D=10$ with one gauge group, $SO_{32}$, occurring with both possible SUSYs.
  The corresponding results for $0$ to $6$ compactifications is presented in \autoref{table:models}.

  \begin{table}[!ht]
    \setlength{\extrarowheight}{.5ex}
    \begin{tabular}{l|c|c|c|c|c|c|c}
      \hline
                     & $D=10$ & $D=9$ & $D=8$ & $D=7$ & $D=6$ & $D=5$ & $D=4$ \\\hline
      $\N = \NMax$   & $2$    & $9$   & $13$  & $16$  & $18$  & $40$  & $68$  \\
      $\NZero$       & $6$    & $32$  & $50$  & $85$  & $73$  & $292$ & $509$ \\
      \textbf{Both}  & $1$    & $3$   & $6$   & $8$   & $18$  & $26$  & $50$  \\\hline
    \end{tabular}
    \caption{
      \textbf{Number of Unique Models} -
      Number of unique \mbox{$\N = \NMax$} and $\NZero$ models for each value of $D$.
      Also included is the number of models that have both $\N = \NMax$ and $\NZero$ realizations.
    }\label{table:models}
  \end{table}

  The general trend, an increase in unique models with compactification, is expected; each compactification adds an addition $U_1$ gauge degree of freedom.
  In the WCFFHS formalism, several factors influence how the $U_1$ alters the initial gauge group: it may enhance the initial gauge group, manifest as an additional group factor, or in more rare cases result in a splitting of group factors.

  Upon construction of all unique models it is a straight forward matter to consider the rate of occurrence of various combinations of group factors which is presented in the tables in the appendix.
  Of particular interest is the emergence of GUT groups with compactification.
  As most common GUTs require multiple low-rank special unitary group factors, a notable exception being SUSY \SO{10}, they arise significantly more often in low dimensions.
  This can be attributed to the enhancement of the additional \U{1} factors produced with compactification.
  Additionally, the application of the GSO projection to reduce the number of spacetime supersymmetries has the tendency of increasing the production of special unitary groups.
  This leads to {\NZero} models favoring the occurrence of the ``unitary GUTs" while the {\NMax} models favor those with special orthogonal groups.

  Beyond GUT groups we can examine how individual group factors arise.
  As an example, consider \SU{3} whose first occurrence is at $D=7$, {\NZero} and first manifests with {\NMax} at $D=5$.
  In all cases, \SU{3} occurs in conjunction with a \U{1} factor which gives rise to the MSSM group in any situation where \SU{2} and \SU{3} arise together.
  Of particular interest is the manner in which \SU{3} is produced.
  Specifically, each compactification produces a \U{1} gauge charge.
  If this charge does not enhance any of the present factors it may remain external, in which case a subsequent compactification is likely to provided an enhancement to $\SU{2} \otimes \SU{2}$.
  A further compactification often yields another enhancement to \SU{4} suggesting that the \SU{3} does not result from the typical enhancement pattern.
  In fact, it is believed that the most probable method of producing an \SU{3} is via breaking of the \SU{4} previously described. This is due to the combination of the additional \U{1} charge and the application of the GSO projection to reduce the model to {\NZero}.
  Additional considerations of this analysis style will be carried out and presented in a future work; however, this represents a departure from the conventional method of analysis of WCFFHS models in that consideration is given to the affect of compactification on a single model rather than considering only models in a fixed dimension.

    \section{Conclusion}\label{conclusion}
  In this survey we can see how the structure of the landscape depends upon compactification.
  Much of the richness of the theory arises primarily from the act of compactification due to addition of \U{1} gauge degrees of freedom.
  Of particular interest here is the occurrence of GUT groups; compactification significantly increases the occurrence of low-rank special unitary groups and thereby common GUT groups.
  It is not until we compactify $5$ dimensions that the $\SM$ arises.
  A deeper look at the evolution of models upon compactification will be studied in a future publication.
    \onecolumngrid
\appendix*
  \section{Gauge Group Combinations}\label{appendix}
    Herein we present statistics for occurrence of specific group factors in various combinations across the layer $1$ landscape.
    As well as combinations of two group factors, we look at combinations of specific compound factors in conjunction with single and other compound factors.
    Following \cite{Moore:2011} we include such compound factors as $\EE{6}{6}$, $\G{PS} \equiv \PS$ (Pati-Salam), $\G{LRS} \equiv \LRS$ (Left-Right Symmetric), and $\G{SM} \equiv \SM$ (Standard Model).
    We also include $\FSUfive \equiv \FSU{5}$, though, because we are not considering matter content, we can only say that the model has the $\FSUfive$ gauge group; it may not actually be $\FSUfive$.

    The percentage of all unique $\NMax$ and $\NZero$ models in $D=10$ through $D=4$ with each combination of gauge groups is tabulated.
    As an example, $11.76\%$ of the $68$ unique $\N = 4$ models, \autoref{subtable:D4_NMax}, have the combination $\SU{4} \otimes \U{1}$ at least once.

    \begin{table*}[!ht]
      \begin{subtable}[b]{.3\textwidth}
        \begin{tabular}{|c|c|c|}
          \hline
          $N=2$       & \SON  & \E{N} \\\hline\hline
          \SON        & 0     & 0     \\\hline
          \E{N}       & --    & 50.00 \\\hline\hline
          Total       & 50.00 & 50.00 \\\hline
        \end{tabular}
        \caption{\NMax}
        \label{subtable:D10_NMax}
      \end{subtable}
      \quad
      \begin{subtable}[b]{.45\textwidth}
        \begin{tabular}{|c|c|c|c|c|c|c|}
          \hline
          $N=6$       & \U{1} & \SU{2} & \SUN  & \SO{8} & \SON  & \E{N} \\\hline\hline
          \U{1}       & 0     & 0      & 16.67 & 0      & 0     & 0     \\\hline
          \SU{2}      & --    & 16.67  & 0     & 0      & 0     & 16.67 \\\hline
          \SUN        & --    & --     & 0     & 0      & 0     & 0     \\\hline
          \SO{8}      & --    & --     & --    & 0      & 16.67 & 0     \\\hline
          \SO{10}     & --    & --     & --    & --     & 0     & 0     \\\hline
          \SON        & --    & --     & --    & --     & 16.67 & 16.67 \\\hline
          \E{N}       & --    & --     & --    & --     & --    & 16.67 \\\hline\hline
          Total       & 16.67 & 16.67  & 16.67 & 16.67  & 66.67 & 33.33 \\\hline
        \end{tabular}
        \caption{\NZero}
        \label{subtable:D10_NZero}
      \end{subtable}
      \caption{$D=10$}
      \label{table:D10}
    \end{table*}

    \begin{table*}[!ht]
      \begin{subtable}[b]{\textwidth}
        \begin{tabular}{|c|c|c|c|c|c|c|c|}
          \hline
          $N=9$       & \U{1} & \SU{2} & \SU{4} & \SUN  & \SO{10} & \SON  & \E{N} \\\hline\hline
          \U{1}       & 0     & 0      & 0      & 0     & 0       & 22.22 & 11.11 \\\hline
          \SU{2}      & --    & 11.11  & 0      & 11.11 & 0       & 0     & 0     \\\hline
          \SU{4}      & --    & --     & 0      & 0     & 0       & 0     & 11.11 \\\hline
          \SUN        & --    & --     & --     & 0     & 0       & 0     & 11.11 \\\hline
          \SO{10}     & --    & --     & --     & --    & 0       & 11.11 & 0     \\\hline
          \SON        & --    & --     & --     & --    & --      & 11.11 & 11.11 \\\hline
          \E{N}       & --    & --     & --     & --    & --      & --    & 22.22 \\\hline\hline
          Total       & 33.33 & 11.11  & 11.11  & 22.22 & 11.11   & 55.56 & 44.44 \\\hline
        \end{tabular}
        \caption{\NMax}
        \label{subtable:D9_NMax}
      \end{subtable}
      \quad
      \begin{subtable}[b]{\textwidth}
        \begin{tabular}{|c|c|c|c|c|c|c|c|c||c|c|}
          \hline
          $N=32$      & \U{1} & \SU{2} & \SU{4} & \SUN  & \SO{8} & \SO{10} & \SON  & \E{N} & \EE{6}{6} & \G{PS} \\\hline\hline
          \U{1}       & 21.88 & 28.13  & 12.50  & 40.63 & 12.50  & 6.25    & 25.00 & 15.63 & 3.13      & 3.13   \\\hline
          \SU{2}      & --    & 34.38  & 12.50  & 15.63 & 0      & 0       & 18.75 & 15.63 & 3.13      & 3.13   \\\hline
          \SU{4}      & --    & --     & 0      & 12.50 & 0      & 3.13    & 9.38  & 6.25  & 0         & 0      \\\hline
          \SUN        & --    & --     & --     & 12.50 & 3.13   & 6.25    & 3.13  & 3.13  & 0         & 3.13   \\\hline
          \SO{8}      & --    & --     & --     & --    & 6.25   & 6.25    & 12.50 & 3.13  & 0         & 0      \\\hline
          \SO{10}     & --    & --     & --     & --    & --     & 0       & 6.25  & 3.13  & 0         & 0      \\\hline
          \SON        & --    & --     & --     & --    & --     & --      & 12.50 & 12.50 & 0         & 6.25   \\\hline
          \E{N}       & --    & --     & --     & --    & --     & --      & --    & 6.25  & 0         & 0      \\\hline\hline
          \EE{6}{6}   & --    & --     & --     & --    & --     & --      & --    & --    & 0         & 0      \\\hline
          \G{PS}      & --    & --     & --     & --    & --     & --      & --    & --    & --        & 0      \\\hline\hline
          Total       & 71.88 & 40.63  & 21.88  & 40.63 & 21.88  & 15.63   & 50.00 & 25.00 & 3.13      & 9.38   \\\hline
        \end{tabular}
        \caption{\NZero}
        \label{subtable:D9_NZero}
      \end{subtable}
      \caption{$D=9$}
      \label{table:D9}
    \end{table*}

    \begin{table*}[!ht]
      \begin{subtable}[b]{\textwidth}
        \begin{tabular}{|c|c|c|c|c|c|c|c|}
          \hline
          $N=13$      & \U{1} & \SU{2} & \SU{4} & \SUN  & \SO{8} & \SON  & \E{N} \\\hline\hline
          \U{1}       & 0     & 0      & 0      & 7.69  & 0      & 0     & 7.69  \\\hline
          \SU{2}      & --    & 23.08  & 0      & 7.69  & 0      & 23.08 & 15.38 \\\hline
          \SU{4}      & --    & --     & 0      & 7.69  & 0      & 0     & 0     \\\hline
          \SUN        & --    & --     & --     & 7.69  & 0      & 0     & 7.69  \\\hline
          \SO{8}      & --    & --     & --     & --    & 0      & 0     & 7.69  \\\hline
          \SON        & --    & --     & --     & --    & --     & 23.08 & 15.38 \\\hline
          \E{N}       & --    & --     & --     & --    & --     & --    & 15.38 \\\hline\hline
          Total       & 7.69  & 38.46  & 7.69   & 30.77 & 7.69   & 53.85 & 38.46 \\\hline
        \end{tabular}
        \caption{\NMax}
        \label{subtable:D8_NMax}
      \end{subtable}

      \begin{subtable}[b]{\textwidth}
        \begin{tabular}{|c|c|c|c|c|c|c|c|c||c|c|}
          \hline
          $N=50$      & \U{1} & \SU{2} & \SU{4} & \SUN  & \SO{8} & \SO{10} & \SON   & \E{N} & \EE{6}{6} & \G{PS} \\\hline\hline
          \U{1}       & 32.00 & 24.00  & 20.00  & 44.00 & 0      & 8.00    & 2.00   & 8.00  & 2.00      & 2.00   \\\hline
          \SU{2}      & --    & 46.00  & 6.00   & 24.00 & 16.00  & 2.00    & 32.00  & 16.00 & 0         & 0      \\\hline
          \SU{4}      & --    & --     & 4.00   & 18.00 & 0      & 2.00    & 0      & 4.00  & 2.00      & 0      \\\hline
          \SUN        & --    & --     & --     & 22.00 & 0      & 8.00    & 2.00   & 6.00  & 0         & 2.00   \\\hline
          \SO{8}      & --    & --     & --     & --    & 8.00   & 0       & 18.00  & 6.00  & 0         & 0      \\\hline
          \SO{10}     & --    & --     & --     & --    & --     & 0       & 0      & 0     & 0         & 0      \\\hline
          \SON        & --    & --     & --     & --    & --     & --      & 18.00  & 12.00 & 0         & 0      \\\hline
          \E{N}       & --    & --     & --     & --    & --     & --      & --     & 8.00  & 0         & 0      \\\hline\hline
          \EE{6}{6}   & --    & --     & --     & --    & --     & --      & --     & --    & 0         & 0      \\\hline
          \G{PS}      & --    & --     & --     & --    & --     & --      & --     & --    & --        & 0      \\\hline\hline
          Total       & 46.00 & 64.00  & 20.00  & 44.00 & 24.00  & 8.00    & 46.00  & 26.00 & 2.00      & 2.00   \\\hline
        \end{tabular}
        \caption{\NZero}
        \label{subtable:D8_NMZero}
      \end{subtable}
      \caption{$D=8$}
      \label{table:D8}
    \end{table*}

    \begin{table*}[!ht]
      \begin{subtable}[b]{\textwidth}
        \begin{tabular}{|c|c|c|c|c|c|c|c|c||c|}
          \hline
          $N=16$      & \U{1} & \SU{2} & \SU{4} & \SUN  & \SO{8} & \SO{10} & \SON  & \E{N} & \EE{6}{6} \\\hline\hline
          \U{1}       & 0     & 12.50  & 0      & 18.75 & 0      & 0       & 18.75 & 12.50 & 6.25      \\\hline
          \SU{2}      & --    & 6.25   & 0      & 12.50 & 0      & 0       & 6.25  & 12.50 & 0         \\\hline
          \SU{4}      & --    & --     & 0      & 0     & 0      & 0       & 12.50 & 6.25  & 0         \\\hline
          \SUN        & --    & --     & --     & 12.50 & 6.25   & 6.25    & 6.25  & 6.25  & 0         \\\hline
          \SO{8}      & --    & --     & --     & --    & 0      & 0       & 0     & 0     & 0         \\\hline
          \SO{10}     & --    & --     & --     & --    & --     & 0       & 0     & 6.25  & 0         \\\hline
          \SON        & --    & --     & --     & --    & --     & --      & 18.75 & 12.50 & 0         \\\hline
          \E{N}       & --    & --     & --     & --    & --     & --      & --    & 18.75 & 6.25      \\\hline\hline
          \EE{6}{6}   & --    & --     & --     & --    & --     & --      & --    & --    & 0         \\\hline\hline
          Total       & 37.50 & 18.75  & 18.75  & 37.50 & 6.25   & 12.50   & 50.00 & 37.50 & 6.25      \\\hline
        \end{tabular}
        \caption{\NMax}
        \label{subtable:D7_NMax}
      \end{subtable}

      \begin{subtable}[b]{\textwidth}
        \begin{tabular}{|c|c|c|c|c|c|c|c|c|c|c||c|c|c|}
          \hline
          $N=85$      & \U{1} & \SU{2} & \SU{3} & \SU{4} & \SU{5} & \SUN  & \SO{8} & \SO{10} & \SON  & \E{N} & \EE{6}{6} & \FSUfive & \G{PS} \\\hline\hline
          \U{1}       & 45.88 & 38.82  & 4.71   & 20.00  & 3.53   & 51.76 & 12.94  & 12.94   & 18.82 & 17.65 & 2.35      & 3.53     & 4.71   \\\hline
          \SU{2}      & --    & 36.47  & 0      & 10.59  & 0      & 28.24 & 5.88   & 9.41    & 18.82 & 14.12 & 0         & 0        & 4.71   \\\hline
          \SU{3}      & --    & --     & 1.18   & 0      & 2.35   & 3.53  & 0      & 0       & 0     & 0     & 0         & 2.35     & 0      \\\hline
          \SU{4}      & --    & --     & --     & 8.24   & 0      & 17.65 & 7.06   & 3.53    & 9.41  & 4.71  & 0         & 0        & 1.18   \\\hline
          \SU{5}      & --    & --     & --     & --     & 2.35   & 2.35  & 0      & 0       & 0     & 0     & 0         & 2.35     & 0      \\\hline
          \SUN        & --    & --     & --     & --     & --     & 25.88 & 5.88   & 9.41    & 7.06  & 7.06  & 0         & 2.35     & 4.71   \\\hline
          \SO{8}      & --    & --     & --     & --     & --     & --    & 4.71   & 0       & 9.41  & 4.71  & 1.18      & 0        & 0      \\\hline
          \SO{10}     & --    & --     & --     & --     & --     & --    & --     & 2.35    & 3.53  & 3.53  & 1.18      & 0        & 1.18   \\\hline
          \SON        & --    & --     & --     & --     & --     & --    & --     & --      & 11.76 & 10.59 & 0         & 0        & 2.35   \\\hline
          \E{N}       & --    & --     & --     & --     & --     & --    & --     & --      & --    & 5.88  & 0         & 0        & 3.53   \\\hline\hline
          \EE{6}{6}   & --    & --     & --     & --     & --     & --    & --     & --      & --    & --    & 0         & 0        & 0      \\\hline
          \FSUfive    & --    & --     & --     & --     & --     & --    & --     & --      & --    & --    & --        & 2.35     & 0      \\\hline
          \G{PS}      & --    & --     & --     & --     & --     & --    & --     & --      & --    & --    & --        & --       & 0      \\\hline
          Total       & 75.29 & 48.24  & 4.71   & 30.59  & 3.53   & 55.29 & 21.18  & 17.65   & 37.65 & 24.71 & 2.35      & 3.53     & 8.24   \\\hline
        \end{tabular}
        \caption{\NZero}
        \label{subtable:D7_NZero}
      \end{subtable}
      \caption{$D=7$}
      \label{table:D7}
    \end{table*}

    \begin{table*}[!ht]
      \begin{subtable}[b]{\textwidth}
        \begin{tabular}{|c|c|c|c|c|c|c|c|c|c|c||c|}
          \hline
          $N=18$      & \U{1} & \SU{2} & \SU{4} & \SUN  & \SO{8} & \SO{10} & \SON  & \E{N} & \EE{6}{6} \\\hline\hline
          \U{1}       & 11.11 & 0      & 0      & 11.11 & 0      & 5.56    & 5.56  & 5.56  & 5.56      \\\hline
          \SU{2}      & --    & 22.22  & 0      & 11.11 & 0      & 0       & 11.11 & 5.56  & 0         \\\hline
          \SU{4}      & --    & --     & 0      & 5.56  & 0      & 0       & 0     & 5.56  & 0         \\\hline
          \SUN        & --    & --     & --     & 16.67 & 0      & 11.11   & 5.56  & 5.56  & 0         \\\hline
          \SO{8}      & --    & --     & --     & --    & 5.56   & 0       & 11.11 & 5.56  & 0         \\\hline
          \SO{10}     & --    & --     & --     & --    & --     & 0       & 0     & 0     & 0         \\\hline
          \SON        & --    & --     & --     & --    & --     & --      & 16.67 & 16.67 & 0         \\\hline
          \E{N}       & --    & --     & --     & --    & --     & --      & --    & 16.67 & 5.56      \\\hline\hline
          \EE{6}{6}   & --    & --     & --     & --    & --     & --      & --    & --    & 0         \\\hline\hline
          Total       & 16.67 & 22.22  & 5.56   & 38.89 & 22.22  & 11.11   & 50.00 & 33.33 & 5.56      \\\hline
        \end{tabular}
        \caption{\NMax}
        \label{subtable:D6_NMax}
      \end{subtable}

      \begin{subtable}[b]{\textwidth}
        \begin{tabular}{|c|c|c|c|c|c|c|c|c|c|c||c|c|c|}
          \hline
          $N=73$      & \U{1} & \SU{2} & \SU{3} & \SU{4} & \SU{5} & \SUN  & \SO{8} & \SO{10} & \SON  & \E{N} & \EE{6}{6} & \FSUfive & \G{PS} \\\hline\hline
          \U{1}       & 45.21 & 23.29  & 6.85   & 20.55  & 5.48   & 45.21 & 8.22   & 12.33   & 6.85  & 10.96 & 2.74      & 5.48     & 8.22   \\\hline
          \SU{2}      & --    & 41.10  & 0      & 9.59   & 0      & 24.66 & 9.59   & 2.74    & 15.07 & 10.96 & 0         & 0        & 5.48   \\\hline
          \SU{3}      & --    & --     & 2.74   & 0      & 4.11   & 4.11  & 0      & 0       & 0     & 0     & 0         & 4.11     & 0      \\\hline
          \SU{4}      & --    & --     & --     & 10.96  & 0      & 17.81 & 4.11   & 4.11    & 1.37  & 2.74  & 0         & 0        & 4.11   \\\hline
          \SU{5}      & --    & --     & --     & --     & 2.74   & 2.74  & 0      & 0       & 0     & 0     & 0         & 2.74     & 0      \\\hline
          \SUN        & --    & --     & --     & --     & --     & 24.66 & 6.85   & 8.22    & 5.48  & 6.85  & 0         & 2.74     & 6.85   \\\hline
          \SO{8}      & --    & --     & --     & --     & --     & --    & 6.85   & 0       & 10.96 & 5.48  & 0         & 0        & 0      \\\hline
          \SO{10}     & --    & --     & --     & --     & --     & --    & --     & 2.74    & 0     & 2.74  & 1.37      & 0        & 0      \\\hline
          \SON        & --    & --     & --     & --     & --     & --    & --     & --      & 9.59  & 10.96 & 0         & 0        & 0      \\\hline
          \E{N}       & --    & --     & --     & --     & --     & --    & --     & --      & --    & 8.22  & 1.37      & 0        & 0      \\\hline\hline
          \EE{6}{6}   & --    & --     & --     & --     & --     & --    & --     & --      & --    & --    & 0         & 0        & 0      \\\hline
          \FSUfive    & --    & --     & --     & --     & --     & --    & --     & --      & --    & --    & --        & 2.74     & 0      \\\hline
          \G{PS}      & --    & --     & --     & --     & --     & --    & --     & --      & --    & --    & --        & --       & 1.37   \\\hline
          Total       & 58.90 & 45.21  & 6.85   & 21.92  & 5.48   & 52.05 & 26.03  & 13.70   & 32.88 & 26.03 & 2.74      & 5.48     & 8.22   \\\hline
        \end{tabular}
        \caption{\NZero}
        \label{subtable:D6_NZero}
      \end{subtable}
      \caption{$D=6$}
      \label{table:D6}
    \end{table*}

    \begin{table*}[!ht]
      \begin{subtable}[b]{\textwidth}
        \begin{tabular}{|c|c|c|c|c|c|c|c|c|c|c||c|c|}
          \hline
          $N=40$      & \U{1} & \SU{2} & \SU{3} & \SU{4} & \SU{5} & \SUN  & \SO{8} & \SO{10} & \SON  & \E{N} & \EE{6}{6} & \FSUfive \\\hline\hline
          \U{1}       & 2.50  & 12.50  & 2.50   & 5.00   & 5.00   & 40.00 & 7.50   & 7.50    & 25.00 & 15.00 & 2.50      & 0        \\\hline
          \SU{2}      & --    & 12.50  & 0      & 5.00   & 0      & 12.50 & 2.50   & 2.50    & 7.50  & 5.00  & 2.50      & 0        \\\hline
          \SU{3}      & --    & --     & 0      & 0      & 0      & 2.50  & 0      & 0       & 0     & 0     & 0         & 0        \\\hline
          \SU{4}      & --    & --     & --     & 2.50   & 0      & 10.00 & 0      & 2.50    & 5.00  & 5.00  & 0         & 0        \\\hline
          \SU{5}      & --    & --     & --     & --     & 2.50   & 2.50  & 0      & 0       & 0     & 0     & 0         & 2.50     \\\hline
          \SUN        & --    & --     & --     & --     & --     & 25.00 & 5.00   & 10.00   & 12.50 & 10.00 & 0         & 2.50     \\\hline
          \SO{8}      & --    & --     & --     & --     & --     & --    & 2.50   & 0       & 2.50  & 2.50  & 0         & 0        \\\hline
          \SO{10}     & --    & --     & --     & --     & --     & --    & --     & 2.50    & 5.00  & 2.50  & 0         & 0        \\\hline
          \SON        & --    & --     & --     & --     & --     & --    & --     & --      & 15.00 & 15.00 & 0         & 0        \\\hline
          \E{N}       & --    & --     & --     & --     & --     & --    & --     & --      & --    & 10.00 & 2.50      & 0        \\\hline\hline
          \EE{6}{6}   & --    & --     & --     & --     & --     & --    & --     & --      & --    & --    & 0         & 0        \\\hline
          \FSUfive    & --    & --     & --     & --     & --     & --    & --     & --      & --    & --    & --        & 0        \\\hline\hline
          Total       & 62.50 & 20.00  & 2.50   & 17.50  & 5.00   & 52.50 & 10.00  & 17.50   & 47.50 & 27.50 & 2.50      & 5.00     \\\hline
        \end{tabular}
        \caption{\NMax}
        \label{subtable:D5_NMax}
      \end{subtable}

      \begin{subtable}[b]{\textwidth}
        \begin{tabular}{|c|c|c|c|c|c|c|c|c|c|c||c|c|c|c|c|}
          \hline
          $N=292$     & \U{1} & \SU{2} & \SU{3} & \SU{4} & \SU{5} & \SUN  & \SO{8} & \SO{10} & \SON  & \E{N} & \EE{6}{6} & \FSUfive & \G{PS} & \G{LRS} & \G{SM} \\\hline\hline
          \U{1}       & 68.15 & 47.60  & 15.07  & 35.96  & 15.75  & 63.36 & 16.10  & 13.70   & 18.15 & 14.04 & 1.71      & 15.75    & 13.70  & 4.45    & 7.53   \\\hline
          \SU{2}      & --    & 41.78  & 7.53   & 22.95  & 6.85   & 36.99 & 9.59   & 7.88    & 13.36 & 8.90  & 0.68      & 6.85     & 9.25   & 1.03    & 4.45   \\\hline
          \SU{3}      & --    & --     & 9.25   & 4.79   & 5.82   & 10.27 & 1.03   & 1.37    & 0     & 1.03  & 0         & 5.82     & 2.05   & 2.74    & 4.45   \\\hline
          \SU{4}      & --    & --     & --     & 16.44  & 6.51   & 26.03 & 7.53   & 5.14    & 7.19  & 4.45  & 0         & 6.51     & 7.19   & 2.05    & 3.42   \\\hline
          \SU{5}      & --    & --     & --     & --     & 6.51   & 10.62 & 0.68   & 0.68    & 0     & 0.34  & 0         & 6.51     & 1.71   & 2.05    & 2.74   \\\hline
          \SUN        & --    & --     & --     & --     & --     & 29.79 & 9.59   & 9.59    & 10.27 & 8.22  & 0.34      & 10.62    & 8.90   & 1.71    & 4.79   \\\hline
          \SO{8}      & --    & --     & --     & --     & --     & --    & 4.11   & 2.40    & 5.82  & 3.42  & 0         & 0.68     & 3.77   & 0       & 0      \\\hline
          \SO{10}     & --    & --     & --     & --     & --     & --    & --     & 2.40    & 3.42  & 2.74  & 0.68      & 0.68     & 1.71   & 0       & 0      \\\hline
          \SON        & --    & --     & --     & --     & --     & --    & --     & --      & 6.85  & 6.85  & 0.34      & 0        & 2.74   & 0       & 0      \\\hline
          \E{N}       & --    & --     & --     & --     & --     & --    & --     & --      & --    & 4.11  & 0.34      & 0.34     & 0.68   & 0       & 0      \\\hline\hline
          \EE{6}{6}   & --    & --     & --     & --     & --     & --    & --     & --      & --    & --    & 0         & 0        & 0      & 0       & 0      \\\hline
          \FSUfive    & --    & --     & --     & --     & --     & --    & --     & --      & --    & --    & --        & 6.51     & 1.71   & 2.05    & 2.74   \\\hline
          \G{PS}      & --    & --     & --     & --     & --     & --    & --     & --      & --    & --    & --        & --       & 2.74   & 0.34    & 0.34   \\\hline
          \G{LRS}     & --    & --     & --     & --     & --     & --    & --     & --      & --    & --    & --        & --       & --     & 1.03    & 1.03   \\\hline
          \G{RSM}     & --    & --     & --     & --     & --     & --    & --     & --      & --    & --    & --        & --       & --     & --      & 2.74   \\\hline
          \G{SM}      & --    & --     & --     & --     & --     & --    & --     & --      & --    & --    & --        & --       & --     & --      & 2.74   \\\hline\hline
          Total       & 87.67 & 54.79  & 15.07  & 40.75  & 15.75  & 65.07 & 20.55  & 17.12   & 26.71 & 18.15 & 1.71      & 15.75    & 16.78  & 4.45    & 7.53   \\\hline
        \end{tabular}
        \caption{\NZero}
        \label{subtable:D5_NZero}
      \end{subtable}
      \caption{$D=5$}
      \label{table:D5}
    \end{table*}

    \begin{table*}[!ht]
      \begin{subtable}[b]{\textwidth}
        \begin{tabular}{|c|c|c|c|c|c|c|c|c|c|c||c|c|c|}
          \hline
          $N=68$      & \U{1} & \SU{2} & \SU{3} & \SU{4} & \SU{5} & \SUN  & \SO{8} & \SO{10} & \SON  & \E{N} & \EE{6}{6} & \FSUfive & \G{PS} \\\hline\hline
          \U{1}       & 29.41 & 11.76  & 4.41   & 11.76  & 5.88   & 36.76 & 4.41   & 7.35    & 11.76 & 5.88  & 1.47      & 5.88     & 2.94   \\\hline
          \SU{2}      & --    & 29.41  & 0      & 7.35   & 0      & 20.59 & 11.76  & 7.35    & 19.12 & 10.29 & 0         & 0        & 1.47   \\\hline
          \SU{3}      & --    & --     & 2.94   & 0      & 1.47   & 1.47  & 0      & 0       & 0     & 0     & 0         & 1.47     & 0      \\\hline
          \SU{4}      & --    & --     & --     & 5.88   & 0      & 11.76 & 1.47   & 4.41    & 1.47  & 2.94  & 1.47      & 0        & 2.94   \\\hline
          \SU{5}      & --    & --     & --     & --     & 4.41   & 2.94  & 0      & 0       & 0     & 0     & 0         & 4.41     & 0      \\\hline
          \SUN        & --    & --     & --     & --     & --     & 33.82 & 7.35   & 13.24   & 13.24 & 8.82  & 1.47      & 2.94     & 4.41   \\\hline
          \SO{8}      & --    & --     & --     & --     & --     & --    & 4.41   & 0       & 5.88  & 1.47  & 0         & 0        & 0      \\\hline
          \SO{10}     & --    & --     & --     & --     & --     & --    & --     & 2.94    & 0     & 2.94  & 0         & 0        & 0      \\\hline
          \SON        & --    & --     & --     & --     & --     & --    & --     & --      & 17.65 & 13.24 & 0         & 0        & 0      \\\hline
          \E{N}       & --    & --     & --     & --     & --     & --    & --     & --      & --    & 8.82  & 1.47      & 0        & 1.47   \\\hline\hline
          \EE{6}{6}   & --    & --     & --     & --     & --     & --    & --     & --      & --    & --    & 0         & 0        & 0      \\\hline
          \FSUfive    & --    & --     & --     & --     & --     & --    & --     & --      & --    & --    & --        & 4.41     & 0      \\\hline
          \G{PS}      & --    & --     & --     & --     & --     & --    & --     & --      & --    & --    & --        & --       & 0      \\\hline\hline
          Total       & 45.59 & 44.12  & 4.41   & 16.18  & 5.88   & 54.41 & 16.18  & 13.24   & 44.12 & 23.53 & 2.94      & 5.88     & 5.88   \\\hline
        \end{tabular}
        \caption{\NMax}
        \label{subtable:D4_NMax}
      \end{subtable}

      \begin{subtable}[b]{\textwidth}
        \begin{tabular}{|c|c|c|c|c|c|c|c|c|c|c||c|c|c|c|c|}
          \hline
          $N=509$     & \U{1} & \SU{2} & \SU{3} & \SU{4} & \SU{5} & \SUN  & \SO{8} & \SO{10} & \SON  & \E{N} & \EE{6}{6} & \FSUfive & \G{PS} & \G{LRS} & \G{SM} \\\hline\hline
          \U{1}       & 75.83 & 50.10  & 25.74  & 40.47  & 20.63  & 62.87 & 11.59  & 13.16   & 10.22 & 10.02 & 1.38      & 20.63    & 15.72  & 9.23    & 15.13  \\\hline
          \SU{2}      & --    & 43.22  & 15.13  & 24.56  & 12.18  & 38.70 & 12.57  & 8.06    & 12.77 & 9.43  & 0.98      & 12.18    & 9.63   & 5.30    & 9.23   \\\hline
          \SU{3}      & --    & --     & 14.54  & 11.98  & 10.61  & 15.52 & 0.98   & 1.18    & 0     & 0.59  & 0         & 10.61    & 3.93   & 7.07    & 9.43   \\\hline
          \SU{4}      & --    & --     & --     & 18.86  & 9.43   & 27.70 & 6.29   & 5.89    & 4.13  & 3.93  & 0.59      & 9.43     & 8.25   & 3.93    & 7.27   \\\hline
          \SU{5}      & --    & --     & --     & --     & 8.64   & 12.18 & 0.79   & 0.98    & 0     & 0.59  & 0         & 8.64     & 2.55   & 4.32    & 7.07   \\\hline
          \SUN        & --    & --     & --     & --     & --     & 30.84 & 9.63   & 10.02   & 8.64  & 7.86  & 0.39      & 12.18    & 9.23   & 2.95    & 7.07   \\\hline
          \SO{8}      & --    & --     & --     & --     & --     & --    & 4.52   & 1.77    & 5.11  & 3.54  & 0         & 0.79     & 2.36   & 0       & 0      \\\hline
          \SO{10}     & --    & --     & --     & --     & --     & --    & --     & 2.55    & 1.18  & 2.16  & 0.39      & 0.98     & 2.36   & 0       & 0      \\\hline
          \SON        & --    & --     & --     & --     & --     & --    & --     & --      & 5.89  & 5.30  & 0.20      & 0        & 0.39   & 0       & 0      \\\hline
          \E{N}       & --    & --     & --     & --     & --     & --    & --     & --      & --    & 3.14  & 0.20      & 0.59     & 1.57   & 0       & 0      \\\hline\hline
          \EE{6}{6}   & --    & --     & --     & --     & --     & --    & --     & --      & --    & --    & 0         & 0        & 0.20   & 0       & 0      \\\hline
          \FSUfive    & --    & --     & --     & --     & --     & --    & --     & --      & --    & --    & --        & 8.64     & 2.55   & 4.32    & 7.07   \\\hline
          \G{PS}      & --    & --     & --     & --     & --     & --    & --     & --      & --    & --    & --        & --       & 2.95   & 0.98    & 2.16   \\\hline
          \G{LRS}     & --    & --     & --     & --     & --     & --    & --     & --      & --    & --    & --        & --       & --     & 2.55    & 4.32   \\\hline
          \G{RSM}     & --    & --     & --     & --     & --     & --    & --     & --      & --    & --    & --        & --       & --     & --      & 7.07   \\\hline
          \G{SM}      & --    & --     & --     & --     & --     & --    & --     & --      & --    & --    & --        & --       & --     & --      & 7.07   \\\hline\hline
          Total       & 83.69 & 62.87  & 25.74  & 41.65  & 20.63  & 65.62 & 19.25  & 13.75   & 21.41 & 15.32 & 1.38      & 20.63    & 16.50  & 9.23    & 15.13  \\\hline
        \end{tabular}
        \caption{\NZero}
        \label{subtable:D4_NZero}
      \end{subtable}
      \caption{$D=4$}
      \label{table:D4}
    \end{table*}
\twocolumngrid

  \clearpage
  \bibliography{main}

\end{document}